\documentclass[12pt,a4paper]{article}
\usepackage{amssymb}
\usepackage{amsmath}

\def\b#1{{\mathbb #1}}
\def\c#1{{\cal #1}}
\def\Dirac{{\raise0.09em\hbox{/}}\kern-0.69em D}
\def\etal{{\it et al.}\ }
\def\kbar{{\mathchar'26\mkern-9muk}}  
\def\ket#1{\vert #1 \rangle}
\def\p{\partial}
\def\t#1{\tilde{#1}}
\def\tr{\mbox{Tr}}

\def\exterior{{{\raise0.2em\hbox{$\scriptstyle\bigwedge$}}{}}}
\def\wm{\mathbin{*}}

\def\k{\kern-.1em\mathbin{,}\kern-.1em}
\def\hk{\kern.12em\raise-1em\hbox{$\hat{\raise1em\hbox{,}}$}\kern.12em}
\newcommand{\initiate}{\setcounter{equation}{0}}
\def\citeyear#1{\cite{#1}}
\begin{document}

\title{Fuzzy Instantons}

\author{Harald Grosse,$\strut^{1}$\
        Marco Maceda,$\strut^{2}$\\[6pt]
        John Madore,$\strut^{2,3}$ Harold Steinacker$\strut^{2,4}$\\[15pt]
        $\strut^{1}$Institut f\"ur Theoretische Physik, Universit\"at Wien\\
        Boltzmanngasse 5, A-1090 Wien \\[5pt]
        $\strut^{2}$Laboratoire de Physique Th\'eorique\\
        Universit\'e de Paris-Sud, B\^atiment 211, F-91405 Orsay \\[5pt]
        $\strut^{3}$Max-Planck-Institut f\"ur Physik\\
        F\"ohringer Ring 6, D-80805 M\"unchen%\\[5pt]
\and    $\strut^{4}$Sektion Physik der Ludwig--Maximilians--Universit\"at 
        M\"unchen\\
        Theresienstr.\ 37, D-80333 M\"unchen}
        %\\[5pt]

\date{}

\maketitle

\begin{abstract}
  We present a series of instanton-like solutions to a matrix model
  which satisfy a self-duality condition and possess an action whose
  value is, to within a fixed constant factor, an integer $l^2$. For
  small values of the dimension $n^2$ of the matrix algebra the
  integer resembles the result of a quantization condition but as
  $n\to\infty$ the ratio $l/n$ can tend to an arbitrary real number
  between zero and one.
\end{abstract}

\vfill
\noindent
LMU-TPW 05/01\\
\noindent
LPT-ORSAY 01-65\\
\noindent
UWTh-Ph-28-2001\\
\noindent
%DRAFT Version CVS:1.36
\newpage

\initiate
\section{Introduction and motivation}

The ideas behind the construction of solitons and instantons can be to
a certain extent extended to noncommutative geometries, at least to
some of them. Within the present context the word soliton refers to a
finite-energy smooth solution with no reference to conserved
quantities; an instanton is a smooth finite-action solution with
euclidean signature, generally involving Yang-Mills fields whose field
equations reduce to self-duality conditions. In ordinary geometry they
are both stable because of a topological obstruction to their decay.
To a certain extent topology now becomes ill-defined and propagators
regular.  Soliton solutions thus loose somewhat their specificity.  It
is nevertheless interesting to study their structure on a fuzzy space.
It was sometime after Dirac presented his magnetic
monopole~\cite{Dir48} solution that it was shown~\cite{tHo74b} to be a
singular limit of a soliton solution. Since the monopole charge is the
first Chern number of a line bundle over the 2-sphere, to obtain a
regular solution this bundle has to be combined with another of equal
and opposite monopole charge. The gauge group has to be enlarged to
$SU_2$ and a Higgs field added so that the topological twist can be
encoded as the winding number of a map of the sphere at infinity in
configuration space into the unit sphere of the $SU_2$ Lie algebra.
The integer in turn can be identified as an element of the second
homotopy of the homogeneous space $SU_2/U_1$. We recall this history
because the monopole suggests one path from simple electromagnetism to
the Yang-Mills-Higgs-Kibble action with an $SU_2$ gauge group and a
Higgs field with values in the adjoint representation. 

Another way~\cite{Mad89c} is the study of electromagnetism on a
noncommutative Kaluza-Klein-type extension of space-time
with~\cite{DubKerMad89b} the matrix algebra $M_2$ as additional
internal structure. If we consider Maxwell theory on this algebra then
the extra matrix factor transforms the Maxwell potential into a
$U_2$-gauge multiplet and a triplet of Higgs scalars.  The
't~Hooft-Polyakov solution can be considered then as a `Maxwell'
solution on the extended algebra $\c{A}$ and in this sense one can say
that with the extra noncommutative factor the core region of the Dirac
monopole has been regularized.  To define a smooth structure we must
introduce a calculus or algebra of forms, a graded differential
algebra $\Omega^*(\c{A})$ over $\c{A}$. 

The theory which we call Maxwell's on space-time can acquire quite a
different aspect when applied~\cite{Mad89c} to more exotic spaces. We
have just seen that the inclusion of a simple matrix factor in the
algebra can radically change the singularity structure of the original
theory.  There are further examples of this, all of which are
constructed by a procedure which resembles the blowing up of
singularities in algebraic geometry except that in the present case a
singular point is regularized not by expanding it into a higher
dimensional variety but rather by surrounding it with a finite `fuzzy'
structure.  An enlightening comparison is with the
description~\cite{Mad82b} of the core regions of planar nematic liquid
crystals. The order-parameter of such a region can `escape into the
third dimension' thereby resolving itself into a smooth configuration
or it can remain, ill-defined, in the plane.

By definition the local `electromagnetic' gauge group of a geometry
with structure algebra $\c{A}$ is the set of unitary elements of the
algebra. If $\c{A} = M_n$ then this yields $U_n$ as `local' gauge
group. A simple example is afforded by the quantum approximations to a
classical spin for large quantum numbers. If $s$ is the spin then the
spin operators lie in $M_n$ with $n=2s+1$. In a way which can be made
precise the algebra $M_n$ tends to the algebra of smooth functions on
the sphere and the group $U_n$ tends to the group $\c{U}_1$ of local
$U_1$ transformations. We shall refer to a manifold possessing a
filtering of its calculus by a sequence of differential algebras over
algebras of matrices as being fuzzy. As a fuzzy version of the space
$\b{R}^3$ in which a monopole lives one can choose the simplest: the
direct sum of all matrix algebras. 

In Section~2 we shall very briefly compare electromagnetism on
ordinary flat $\b{R}^3$ with the same theory on the slightly more
structured `space' which is a Kaluza-Klein extension~\cite{Mad89c}
thereof by the algebra $M_2$ of $2\times 2$ matrices over the complex
numbers. Although it is necessary to introduce the complex numbers to
describe it, the geometry of this latter `2-point' space is real. We
examine also the analogous case of a time-dependent
fuzzy-sphere~\cite{Mad92a,GroMad92}. The `abelian' gauge theory on the
fuzzy sphere has several ground states which have been
called~\cite{DubKerMad89b} `instantons' but which in the present
context should rather be dubbed `solitons'.  In Section~3 we find
instanton-like solutions which tunnel between two of these solitons.
The formalism of the following section could be said to describe
`fuzzy' spherical $D2$-branes~\cite{BanFisSheSus97}, in which case the
instanton would be said to tunnel between two such configurations, one
in which the branes coincide and the other in which they are
completely separated.  In the spherical case the separation is
quantized and the instanton can have a topological interpretation.  An
summary of some results on projectors used is given in the Appendix.

We stress here the role of instantons as mediators between different
stable vacuum sectors of a matrix geometry.  Several other aspects of
instantons have also been carried over~\cite{BaeBalIdrVai00,%
GopMinStr00, GroNek00b, HarKraLarMar00, ConLan01} into various
noncommutative geometries including the fuzzy
sphere~\cite{IsoKimTanWak01, HasKra01, HikNoz01}. Similar
calculations have been carried out~\cite{KraSch01} on the torus.

\initiate
\section{Abelian gauge theory}

In this section we recall very briefly some formulae which will be
useful in studying product structures. Let $\b{P}^1(\b{C}) = S^2$ be
smoothly embedded in $\b{R}^3$. A Dirac monipole will refer to a
principle $U_1$ bundle over $S^2$ or to an associated complex line
bundle $L$ over $\b{P}^1(\b{C})$ or to an extension of the former to a
$U_1$ bundle over $\b{R}^3$, either of which can be completely
classified by a single integer $k$. Expressed algebraically a Dirac
monopole is an irreducible module over the algebra 
$\c{C} (S^2) \subset \c{C} (\b{R}^3)$; a 't~Hooft-Polyakov monopole or
soliton is exactly the same object over the algebra 
$\c{A} = \c{C}(S^2) \otimes M_2$.  More precisely, if one consider
`Maxwell's' theory on the two algebras and look for a solution which
tends in the asymptotic region to the monopole, then if 
$\c{A} = \c{C}(\b{R}^3)$ one obtains the Dirac monopole and if 
$\c{A} = \c{C} (\b{R}^3) \otimes M_2$ the 't~Hooft-Polyakov soliton.
This is an obvious remark which follows directly from the definition
of a differential calculus over a product structure which is given in
the Appendix. It is perhaps the simplest example of the regularizing
capabilities of noncommutative extensions of manifolds with
singularities. 

There are many noncommutative versions of $\b{R}^3$,
all of which necessarily break the inhomogeneous invariance group
$ISO_3$ of $\b{R}^3$. We mention in particular the $q$-deformed
version of Faddeev \etal~\cite{FadResTak89} which has an interesting
quantum group as invariance `group'. Otherwise one can for example
split the space into the product $\b{R}^3 = \b{C} \times \b{R}$ of the
complex line times the real line~\cite{GroNek00b}, a procedure which
respects the subgroup $ISO_2 \times \b{R} \subset ISO_3$. We shall
choose to split space into the product $\b{R}^3 = S^2 \times \b{R}^+$
which respects the subgroup $SO_3 \subset ISO_3$. Although we consider
here exclusively the case of the canonical flat $\b{R}$ the same
reasoning will apply to a curved manifold provided the soliton in
question is of dimension $r$ and placed in a region of typical
Gaussian curvature $K$ such that $r^2 K \ll 1$. We replace the algebra
$\c{C}(\b{R}^3)$ of smooth functions of compact support by the
noncommutative algebra
$$
\c{A} = \bigoplus_1^\infty M_n.
$$
Each $M_n$ describes a fuzzy 2-sphere~\cite{Mad00c} of radius $r_n$
determined by the possible values of the Casimir operator. If we
introduce a length scale $\sqrt \kbar$ then the possible values of $r$ are
given, for large $n$, by the sequence of solutions to the equation 
\begin{equation}
n = \frac{4\pi r^2}{2\pi\kbar}.                            \label{r-kbar}
\end{equation}
This is to be interpreted as meaning that the `2-sphere' of radius
$r$ describes a `space' consisting of $n$ cells of area $2\pi\kbar$.
The interpretation is taken from quantum mechanics. It parallels also
that of a spin or of the Landau-orbit structure of an electron gas in
a constant normal magnetic field. The $n$ `points' are the ground states
of the Landau orbits; the remaining $n(n-1)$ degrees of freedom of
$M_n$ become in the limit delocalized states.~\cite{Mad00c}
Space acquires an onion-like structure with an infinite sequence of
concentric fuzzy spheres at the radii given by the Casimir relation
(\ref{r-kbar}), which can also be written
$$
\mbox{Area}[S^2] = 2\pi\kbar n
$$
as a quantization condition on the area.  A large number of
differential calculi can be put on a matrix algebra, a subset
$\Omega_m(M_n)$ of which can be described~\cite{Mad00c} in terms of
the action of a group $SU_m$ with $m$ belonging to a subset of the
integers less than $n$. We shall restrict our considerations here
to the simplest situation with $m=2 \ll n$.

The differential calculus is a product calculus; on the first factor
it is that of the fuzzy sphere and along the generator $r$ the
ordinary one in the finite-difference approximation.  In this example
the electromagnetic potential can be thought of as the Higgs field.
Its vacuum value is the `Dirac operator' $\theta$ and it tends at
infinity to the Dirac-monopole solution of change $-1$.  There should
be no obstruction to considering higher monopole
charge~\cite{GroKliPre97a}.  A sequence of projective modules
describing in the limit the higher monopole charge configurations has
been constructed\cite{GroRupStr01}. The Chern numbers which arise
there are noninteger and converge in the commutative limit to the
integer Chern numbers of the complex line bundles over the two-sphere.
To describe the Schwarzschild instanton in terms of matrices we must
replace the factor $M_2$ in $\c{A}$ by $M_l$ with $l\gg 2$.  That is,
we must consider the extensions
$$
\c{A} \mapsto \c{A}^\prime \otimes \c{A}^{\prime\prime} = 
\bigoplus_{l,n=1}^\infty (M_l \otimes M_n)
$$
to the tensor product of two copies of $\c{A}$.  The monopoles
become the first two elements in either one of the factor series.

We recall that a differential calculus can be completely defined in
terms of the left and right module structure of the $\c{A}$-module of
1-forms $\Omega^1(\c{A})$.  We shall restrict our attention to the
case where this module is a submodule of a module of rank $n$ which is
free as a left or right module and which possesses a special basis
$\theta^a$, $1\leq a \leq n$, which commutes with the elements $f$ of
the algebra.
\begin{equation}
[f, \theta^a] = 0.                                             \label{fund}
\end{equation}
The differential $d$ is a projection onto $\Omega^1(\c{A})$ of the
image of a map $d$ given by the expression
\begin{equation}
df = e_a f \theta^a = [\lambda_a, f] \theta^a.            \label{defdiff}
\end{equation}
One can rewrite this equation as
\begin{equation}
df = -[\theta,f],                                          \label{extra}
\end{equation}
if one introduces~\cite{Con94} the `Dirac operator'
\begin{equation}
\theta = - \lambda_a \theta^a.                                 \label{dirac}
\end{equation}
The differential is then of the form
\begin{equation}
df = - [\theta, f] = - e_a f \theta^a.
\end{equation}
The differential of the relations $R$ which define the algebra must
vanish and the largest module of 1-forms is obtained by supposing that
these are the only relations in $\Omega^1(\c{A})$.

In the present case the $\lambda_a$ satisfy the commutation relations
$[\lambda_a, \lambda_b] = C^c{}_{ab} \lambda_c$ with 
$C_{abc} = r^{-1}\epsilon_{abc}$ and they have units of inverse
length. The forms $\theta^a$ anti-commute. The form $\theta$ can be
considered as a flat connection:
$$
d\theta + \theta^2 = 0.
$$
The fuzzy 2-solitons are in 1-1 correspondence with the projectors
and their number is given by the (Hardy-Ramanujan) partition function
$p(n)$.

The calculus we use is a 3-dimensional calculus on the Hopf fibration
over the 2-sphere. This is necessary because the 2-sphere is not
parallelizable.  We would like therefore to have an expression for the
integral of a `function' on the fuzzy 2-sphere as integral of a 3-form.
If $\alpha$ is a $d$-form over a $d$-dimensional manifold then
under Hodge duality it corresponds to an element $a$,
$$
\alpha = a\theta^1 \cdots \theta^d 
\longrightarrow \mbox{Vol}(V) a
$$
of the algebra which has scale dimension $L^{-d}$. It seems quite natural
to define the integral of $a$ so that
$$
\int \alpha =\mbox{Vol}(V) n^{-1} \tr (a).
$$
With a second length scaling becomes rather ambiguous.  The
identity~(\ref{r-kbar}) can be considered as a relation between
$\kbar$ and $n$ with $r$ fixed or as a relation between $r$ and $n$
with $\kbar$ fixed. The limits when $n\to\infty$ are respectively the
sphere of radius $r$ and the noncommutative plane. In the case at hand
the volume $V$ is a circle bundle over one of these objects. We would 
like to include the case where the circle and the sphere scale differently.
We introduce therefore a constant $\epsilon$ which is the ratio of the
corresponding radii and scale $r\to n^{\gamma}r$.
The integral is given in terms of the trace as
\begin{equation}
\int \alpha = \pi^2 r \epsilon n^{3\gamma}\kbar \tr (a).         \label{vol}
\end{equation}
When $\gamma = 0$ the radius $r$ remains fixed and
when $\gamma = 1/2$ the length scale $\kbar$ remains fixed.
If we require that the radius of the circle remain fixed under a change
of $r$ we must choose
$$
\epsilon = n^{-\gamma}.
$$
In the abstract theory of integration on noncommutative
algebras~\cite{Con94} there is but one dimension in which the integral
makes sense and all matrix algebras are considered to be of dimension
zero. We shall enlarge the algebra by adding a time variable and
look for time-dependent configurations which tunnel between two static
configurations.  These are referred to as (fuzzy) instantons.

\initiate
\section{Fuzzy Instantons}

Let $\theta^a$ be a real frame~\cite{DubKerMad89a,DimMad96,FioMad98a}
for the differential calculus over the matrix factor and let 
$\theta^0 = dt$ be the standard de~Rham differential along the real
line. The differential $df$ of $f$ can be written as
$$
df = e_a f \theta^a + \dot f \theta^0. \qquad \dot f = \p_t f
$$
Over the algebra $\c{A}$ we introduce the product calculus
described in Section~3. An arbitrary 1-form $\omega$ can be expanded
in the basis $\theta^\alpha = (\theta^0, \theta^a)$. Using the
notation of the Appendix we can write
$\omega = \omega_\alpha \theta^\alpha$ and therefore in terms of 
$\phi = \omega - \theta$ the curvature
$$
\Omega = d\omega + \omega^2 = 
\frac 12 \Omega_{\alpha\beta}\theta^\alpha \theta^\beta
$$
can be written as
$$
\Omega_{0a} = \dot \phi_a + e_a \phi_0 + [\phi_0, \phi_a], \qquad
\Omega_{ab} = [\phi_a, \phi_b] - C^c{}_{ab} \phi_c.
$$
The structure constants $C^c{}_{ab}$ are those of $SU_2$ and for
convenience they contain an extra factor 
$r^{-1}$: $C_{abc} = r^{-1} \epsilon_{abc}$. The covariant derivative
$D_a$ is defined by
$$
D_a \phi_b = [\phi_a, \phi_b] -  
\frac 12 C^{c}{}_{ab} \phi_c.
$$
The equations we use are the analogs
$$
D_\alpha \Omega^{\alpha\beta} = 0,
$$
of the usual Maxwell equations. We recall that we have changed the
`space' rather than the theory; the local gauge group remains the set
of unitary elements of the algebra.  The equations become
$$
D_a \Omega^{a0} = 0, \qquad
D_0 \Omega^{0b} + D_a \Omega^{ab} = 0.
$$
The first equation yields
$$
D_a \Omega^{a0} = [\phi_a,\Omega^{a0}] = 
- [\phi^a,\dot \phi_a] + [\phi^a, e_a \phi_0] + [\phi^a, [\phi_a, \phi_0]].
$$
If we look for solutions with $\phi_0 = 0$ then we see that
$[\phi^a,\dot \phi_a] = 0$ which implies that 
$\dot \phi \propto \phi$.  If we choose the Coulombe gauge there is
one particularly simple Ansatz given by $\phi_a = e(t) \lambda_a$.
This can be also written as
$$
\phi = - e(t) \theta, \qquad \omega = (1-e(t))\theta.
$$
The form of the Ansatz is not gauge-invariant. The $e(t)$ is {\it a priori}
an arbitrary element of the algebra. 

The field strength is
\begin{equation}
\Omega = e(\theta e - \theta) \theta + \dot e \theta \theta^0.   \label{curv}
\end{equation}
Since $\theta$ is the `Dirac' operator this is the same as the usual
expression 
$$
\Omega = e de de
$$
when $e$ is a projector, that is in the present context, when 
$t \to \pm\infty$.  We shall compute only the simple case with $e(t)$
in the center.  Using the relation~(\ref{r-kbar}) we find that
the (euclidean) field equations reduce to
$$
r^2 \ddot f - f (f-1)(2f-1) = 0.
$$
The roots $f=0,1$ correspond to the two stable ground states of the
system and the root $f= 1/2$ to the unstable
solution~\cite{DubKerMad89b}. The instanton solution we give below
interpolates between the first two, passing through the third.

An instanton is described by two partitions of an arbitrary integer
$n$, one which characterizes the soliton at $t \rightarrow -\infty$
and the other at $t \rightarrow +\infty$. In the case with $e$ in the
center the condition~(\ref{duality}) simplifies to the differential
equation
\begin{equation}
r^2\dot f^2 = f^2 (f-1)^2,                            \label{fe}
\end{equation}
a nonlinear first-order equation, which is also a first integral of the
field equations. The general solution
\begin{equation}
f = \frac 12 (1 + \tanh (\frac{t}{2r} + b)),  
\qquad b \in \b{R}                                           \label{t-sol}
\end{equation}
interpolates between $f(-\infty) = 0$ and $f(+\infty) = 1$. This has
the same form as the classical double-well instanton~\cite{Pol77}.

We define the integral as usual in terms of the trace and use the
definitions of Sections~2 to define the action as an integral of a
4-form. We set, that is
$$
S = \frac 12 \int_{S^3 \times \b{R}} \Omega \wm \Omega.
$$
The product is in the algebra of forms; the star refers to duality.
As in the commutative case we use the duality
condition~(\ref{duality}) to write the action (formally) proportional
to the 2nd-Chern number:
$$
S = \pm \frac 12  \int_{S^3 \times \b{R}} \Omega^2.
$$
To calculate $S$ however it is more convenient to use the fact that
the 4-form $\tr (\Omega^2)$ is the exterior derivative of a 3-form,
$$
\int_{S^3 \times \b{R}} \Omega^2 = 
\int_{S^3 \times \b{R}} dK, \qquad 
K = \omega d\omega + \frac 23 \omega^3.
$$
We can express then the action
$$
S = \frac 12 \int_{S^3} K(+\infty) - 
\frac 12 \int_{S^3}K(-\infty)
$$
as an ordinary volume integral.  We apply Stokes' theorem only along the
(ordinary) time axes; in general it makes less sense as a relation
amongst traces of operators.  Our solution tunnels between $f=0$ at 
$t =-\infty$ to $f=1$ at $t=\infty$. That is $K(+\infty) = 0$ and
$$
K(-\infty) = - \frac 1{3} \int \theta^3.
$$
But
$$
\int \theta^3 = 
-\int \lambda_a \lambda_b \lambda_c \theta^a \theta^b \theta^c =
\frac 14 \pi^2 \epsilon n^{2+3\gamma}.
$$
We find then that the action of the instanton solution~(\ref{t-sol})
in the case of most interest, with $\epsilon = 1$ and $\gamma = 0$, is
$$
S[\gamma = 0] = \frac 1{3!}\int_{S^3} \theta^3 = 
\frac 1{4!} \pi^2 n^2.
$$
It is singular in the limit when $n\to\infty$. It would remain finite if we
let $\epsilon$ scale as
$$
\epsilon = n^{-2-3\gamma}
$$

In the case of the instanton $T^n_l$ which tunnels
$$
[l, \underset{n-l}{\underbrace{1, \cdots 1}}] 
\buildrel T^n_l \over \longrightarrow [\underset{n}{\underbrace{1, \cdots 1}}]
$$
the action (when $\gamma = 0$) is given by
\begin{equation}
S[T^n_l] = \frac 1{3!}\int_{S^3} \theta^3 = \frac 1{4!} \pi^2 l^2.\label{Bohr}
\end{equation}
This is the basic relation which we find.  It involves a new integer
$l$ which in the commutative limit can combine with $n$ to form an
irrational number, the limit of the sequence $l/n$.

We use in an essential way the fact that the differential calculus
over the matrices is based on a module of 1-forms which is free of
rank 3. With the addition of an euclidean time variable this yields a
differential calculus of rank 4. We can thus introduce a duality
$$
\wm (\theta^a \theta^0) = 
\frac 1{2} g^{ad}\epsilon_{bcd} \theta^b \theta^c, \qquad
\wm (\theta^a \theta^b) =  g^{ac}g^{bd} \epsilon_{cde} \theta^e \theta^0
$$
and look for the special set of solutions for $e(t)$ which satisfy
the self-duality conditions
\begin{equation}
\Omega = \pm \wm \Omega.                                     \label{duality}
\end{equation}
This clarifies the duality in the fuzzy spheres found
previously~\cite{GroMadSte00a} and could also shed light on the
appearance of Chern-Simons terms in the work or Alekseev
\etal~\cite{AleRecSch99}.

In terms of components the self-duality condition can be written
$$
\Omega_{ab} = g^{cd}\epsilon_{abd} \Omega_{c0}
$$
and therefore the self-dual and anti-self-dual parts of the curvature
2-form can be written
$$
\Omega^{\pm} = \frac 12  \Omega^{\pm}_{ab} \theta^a \theta^b, \qquad
\Omega^{\pm}_{ab} = 
\frac 12 (\Omega_{ab} \pm g^{cd}\epsilon_{abd} \Omega_{c0}).
$$
In the projector case these become
$$
\Omega^{\pm}_{ab} = [e\lambda_a, e\lambda_b] - C^c{}_{ab} e\lambda_c
\pm  r C^c{}_{ab} \dot e \lambda_c.
$$
If we raise the indices of the components and introduce the 3-dual
$$
\Omega^c = \frac 12 g^{cd}\epsilon_{abd}\Omega^{ab}
$$
then the duality condition can be written
\begin{equation}
 \dot e \lambda^c = \Omega^c [ e\lambda_a ].        \label{d-omega}
\end{equation}
As usual, from the inequality
$$
\int (\Omega \pm \wm \Omega) \wm (\Omega \pm \wm \Omega) \geq 0
$$
we can conclude that the action $S[\omega]$ of any configuration
$\omega$ `in the same sector' is bounded below by $S[T^n_n]$:
$$
\vert S[\omega] \vert \geq S[T^n_n].
$$
The integer $n$ measures the `amplitude' of the instanton and $l$
measures the extension. In a conformal-invariant theory the value of
the action would not depend on these two parameters.

\initiate
\section{Fuzzy Solitons}

Because the radius of the fuzzy sphere is related to a Casimir
operator and lies in the center of the algebra one can consider field
configurations which are concentrated entirely on one sphere. These we
refer to as (fuzzy) 2-solitons, with the alias `fuzzy $D2$-branes'.
Implicit in the expression for the energy density
$$
\epsilon(r) = \frac 12 \Omega_{\alpha\beta}\Omega^{\alpha\beta}(r)
$$
is a choice of differential calculus along the radial direction
$r=x^4$. The Greek indices take the values $1 \dots 4$. If we embed
the solution in the continuous line and use the radial parameter $r$
instead of euclidean time in the solution we gave in the previous
section we find a (fuzzy) 3-soliton, with the alias `fuzzy
$D3$-brane'. As we shall see in
the next section the equilibrium configuration is a distribution which
satisfies one of the `duality' conditions (\ref{d-omega}). It
is given, for large $n$, by the function
$$
f(l) = \frac 12 (1 - \tanh (2\pi \kbar^{-1} r^2)) =
\frac 12 (1 - \tanh (\pi l)), \qquad l\leq n,
$$
the same as (\ref{t-sol}) below but with the replacement
$$
- t  \mapsto \pi \kbar^{-1} r^3
$$
and with $b = 0$. The solution has a maximum at the origin and drops
off exponentially toward infinity.  

To write an expression for the energy of the $l$th sphere we use the
definition of integral given above.  For large $n$ then and for all 
$l \lesssim n$ we set
$$
E^n_l = \int_{r_l}^{r_{l+1}}\epsilon(r).
$$
If $\gamma=0$ we see that $E^n_l = E_l$ is independent of $n$.
The sum
$$
E^n = \sum_0^n E^n_l
$$
is the total energy.

Classically all the vacuum modes have energy equal to zero; including
the zero-point fluctuations this will no longer be the case. For
example the completely reducible and the irreducible configuration
will have vacuum energy given respectively by
$$
E_1 \simeq \frac 12 \hbar \omega \cdot n^2,
\qquad E_n \simeq \frac 12 \hbar\omega \cdot \sum 2l(2l+1),
\qquad \omega = \frac{\sqrt 2}{r}.
$$
In the first there are $n^2$ modes of equal frequency $\omega$; in
the second, for each $l \lesssim n$ there are $2(2l+1)$ modes of
frequency $l\omega$.  The remaining vacuum energies lie between these
two values.  To discuss this it is convenient to introduce a Fock
space.

\initiate
\section{Fock space}

We have found an instanton solution which tunnels between $f=0$ and
$f=1$. If we return to the definition of the curvature as a functional
of the fields $\phi$ we see that this corresponds to a transition from
the irreducible representation of dimension $n$ to the completely
reducible representation of the same dimension. The former corresponds
to the partition $[n]$ of $n$; the latter to the partition 
$[1, \dots 1]$.  Let $[n_1, \cdots n_k]$ be an arbitrary partition of
$n$, a set of non-increasing integers whose sum is equal to $n$. To it
corresponds a representation of $SO_3$ which is the sum of irreducible
representations of dimension $n_i$. By the same construction there is,
for each index $i$, an instanton which tunnels between $[n_i]$ and
$[1, \dots 1]$.  There are perhaps other instantons, those which
correspond to transitions between non-trivial projectors. If 
$[n_1, \cdots, n_k]$ is a partition of $n$ into $k$ integers then the
corresponding representation is block diagonal and reducible to $k$
representations of dimensions $n_i^2-1$ for $1\le i \le k$. Let $e_i$
be the projector onto the $i$th sector. An arbitrary projector $e$ can
be written in the form
$$
e = \sum_i \epsilon_i e_i, \qquad \epsilon_i = 0,1.
$$
The corresponding expression (\ref{curv}) for the field strength 
is given by
$$
\Omega = \sum_i \Omega_i
$$
with each $\Omega_i$ of the form (\ref{curv}). Each corresponding
$f_i$ evolves independently according to its field equation. The
instantons tunnel therefore between different partitions. Since we
have found no others we shall suppose that~(\ref{t-sol}) is the only
type of instanton. It is the only $SO_3$-invariant type.

If one quantizes the system one obtains a bosonic Fock space of
ordinary `vacuum modes'. Each of the $p(n)$ minima of the potential
gives rise to a tower of states in general different from each other.
Besides these modes there is also a Fock space $\c{F}$ of `tunneling
modes', each with only one quantum number, an integer $l$. We set
$$
\ket{\cdots n_{l_1}, \cdots n_{l_k}, \cdots} = [l_1, \cdots l_k].
$$
The integer $n_{l_j}$ in the $l_j$th position indicates the
presence of $n_{l_j}$ tunneling modes with quantum number $l_j$, each
of which is an irreducible representation of rank $l_j$. The tunneling
modes interact with all the vacuum modes and change their energy
eigenvalues in a rather complicated way.  Without the tunneling modes
the vacuum modes do not interact so we consider the tunneling modes as
responsible for the dynamics. They have a single quantum number, the
integer $l$. An instanton gas is an ensemble of tunneling modes,
considered as a Bose gas. The integer $n$ cannot be considered as
conserved and we can suppose then that the chemical potential
vanishes.  The best description of the dynamics is through some
examples.
\begin{enumerate}
\item In Fock-space notation the basic transition is of the general form
$$
\ket{0,\cdots n_{l_i}, \cdots 0, \cdots} 
\buildrel T^n_{l_i} \over \longleftrightarrow 
\ket{l_i, \cdots\underset{l_i}{\underbrace{0, \cdots 0}}, \cdots 0}.
$$

\item More generally we have
$$
\ket{0, \cdots n_{l_1}-1,\cdots n_{l_i}, \cdots n_{l_k}, \cdots} 
\buildrel T^n_{l_i} \over \longleftrightarrow 
\ket{l_i,  \cdots\underset{l_i}{\underbrace{0, \cdots 0}}, \cdots 
n_{l_k}, \cdots}.
$$
One of the $n_{l_i}$ irreducible components of dimension $l_j$ in the given
representation `decays' into $l_j$ representations of dimension
zero, or the inverse. The probability of this transition
is proportional to the barrier penetration rate
$$
p[T^n_{l_j}] = A[T^n_{l_j}] e^{-S[T^n_{l_j}]}.
$$  
The $A[T^n_{l_i}]$ is a $WKB$ amplitude difficult to
calculate in general. 

\item The transition whereby $n$ `decays' into $n-1$ and $1$ can be
written in the Fock-space notation
$$
\ket{0, \cdots 0, \cdots 1_n}
\buildrel T^n_{n-1} \over \longleftrightarrow
\ket{1, \cdots 0, \cdots 1_{n-1}, 0}.
$$
By our assumption that the only transition is that between the
irreducible and completely reducible representation the transition can
only proceed via the intermediate state
\begin{equation}
\ket{0, \cdots 0, \cdots 1_n} 
\buildrel T^n_{n} \over \longleftrightarrow
\ket{n, \cdots 0, \cdots 0} 
\buildrel T^n_{n-1} \over \longleftrightarrow
\ket{1, \cdots 0, \cdots 1_{n-1},0}.                          \label{n-n-1}
\end{equation}

\item All the different partitions are ground states since the action
  of each vanishes at least in the classical approximation. The
  tunneling phenomena would lift this degeneracy and make some
  partitions more favorable.  Consider the case $n=2$ with its two
  partitions $[2]$ and $[1,1]$.  These are not large values of $n$ and
  the `exact' formula must be used, obtained by replacing $n^2$ by
  $n^2-1$.  The two degenerate levels split by an amount proportional
  to the transition probability $p = A e^{-3\pi^2/4}$.
\end{enumerate}

We can now better formulate the problem of the classical limit. We
recall that this limit is singular from several points of view. The
volume form, for example, is closed but not exact in the limit but it
is exact for all finite $n$.  We found that the action of the
transition between the two extreme partitions of $n$ diverges with $n$
but we claim that the limiting bundle has a well-defined Chern number.
In the analogy with statistical physics the euclidean-time world sheet
of each fuzzy surface appears as a probability distribution. The
action is the energy of the configuration considered in one dimension
higher. Since we are working in one dimension lower than the `physical
one' [sic.]  the instantons resemble in all respects solitons. From
the `gas' point of view the penetration probability is to be thought
of as a creation probability from the vacuum, that is, a distribution
probability for the existence of a configuration. By standard
WKB-approximation calculations we have an estimate of the
barrier-pene\-tration probability $p_l$. In terms of an amplitude
$A[T^n_l]$ it is given by~\cite{LanLif97}
$$
p_l = A[T^n_l] e^{-\pi^2 l^2/4}, \qquad l \lesssim n, \quad l \gg 1.
$$
We shall use the language of statistical physics~\cite{LanLif97}
because of the interpretation of multiple paths as an instanton gas.
As a first approximation we shall suppose that the amplitude is
independent of $l$ and we shall write
$$
p_l = e^{\beta F -\pi^2 l^2/4}.
$$
in terms of $\beta F=\log A$. This is the probability distribution
of a classical Bose-Einstein gas with energy spectrum
$$
E_{l} = \frac 12 \pi^2 l^2.
$$
As in ordinary physics a temperature at which the entropy term and the
energy term just cancel could be indicative of a phase transition. One
interesting temperature which must appear also is a maximal or
Hagedorn temperature $T_H$. Noncommutative geometry certainly requires
that $T_H$ exist and further that
$$
T_H^2 \kbar \lesssim 1.
$$
In the recent literature~\cite{Mag98, Sat98} there are calculations
of such temperatures within the context of matrix models. To a certain
extent the expression $\beta = \sqrt\kbar$ could be said to play this role.
From the expression of the statistical distribution it appears that a
configuration with more isolated $D$-branes is more probable. We can
conclude that they repulse due to the instanton exchange. This would
seem to contradict the results of Gao \& Yang~\cite{GaoYan01}.

\initiate
\section{Speculative conclusion} 

We have calculated transition instantons between various vacua of
`Maxwell's theory' on different matrix geometries.  There are two
`mode gases' which one can introduce.  The modes around the classical
vacua, the `momentum' modes, interact with a gas of tunneling modes,
the instantons; otherwise the gas is free.  The Hopf fibration has
Chern number $c_1 = -1$ and for $n$ large the fuzzy sphere is a good
approximation. One can say then that the instanton tunnels between a
good and very bad approximation to the sphere but not between two
distinct topological sectors. If one take the time variable together
with the circle of the fibration then by compactifying one obtains a
second sphere. The space becomes the euclidean Schwarzschild solution,
whose instanton number is the product~\cite{DufMad78} of an electric
monopole charge and a magnetic monopole charge equal to $-1$.  The
electric charge must be identified as a topological invariant of the
matrix factor. This is {\it strictu sensu} neither possible nor
desirable. But the infinite potential barrier associated with a
topological obstruction can be replaced by a very large barrier which
would tend to infinity with the integer $n$.  We imagine the action of
the partition $[n]$ as a quantum of energy for a state of the bosonic
Fock space described above and that the action of the element of the
instanton which tunnels to the classical manifold appears as the
associated free energy.  We refrained from speculating about possible
relations between the limit of the tunneling modes and winding modes.

\initiate
\section*{Acknowledgments}

One of the authors (J.M.) would like to thank L.~Dabrowski for his
hospitality at S.I.S.S.A. where this work initiated and B.~Schmidt
for his hospitality an the A.~Einstein Institut, Golm where it
terminated. He also thanks P.~Tod and D.~Robinson for enlightening comments. 
H.S. would like to thank J.~Pawelczyk for discussions on Chern-Simons 
terms. All benefited from discussion from G.~Landi.

\initiate
\section{Appendix: projectors}

If we consider a manifold $V$ of dimension $d$ as a subset of some
higher-dimensional euclidean space $\b{R}^n$ then the algebra of
functions $\c{A} = \c{C}(V)$ can be defined as a formal algebra in
terms of generators and relations.  The coordinates $x^i$ of $\b{R}^n$
are the generators of an algebra $\t{\c{A}}$ which satisfy only the
commutation relations $[x^i,x^j]= 0$. Within $\t{\c{A}}$ there is an
ideal defined by constraint relations $R_a(x^i)$ which define $V$ as a
submanifold. If $\t{\c{A}}$ is noncommutative then we suppose it to be
still characterized by commutation relations but we shall no longer
suppose that the constraint relations necessarily generate a
non-trivial 2-sided ideal. The projection of $\t{\c{A}}$ onto $\c{A}$
which describes the embedding in the commutative limit will be a
vector space map which to a certain approximation only can be
considered an algebra epimorphism.  If one introduce the moving frame
$dx^i$ on $\b{R}^n$ then the latter acquires the structure of a flat
differential manifold. This defines by the embedding, that is, by the
relations, a moving frame $\theta^\alpha$ on $V$ and a splitting of
the module of sections $\c{A}^n$ of $T^*(\b{R}^n)$ into a direct sum
$$
\c{A}^n = \Omega^1(V) \oplus \c{N}
$$
of the module of 1-forms and a complement.  The metric and the
frame on $V$ are determined by the embedding relations.  The
construction is most elegantly described in terms of projectors.
Similar constructions on ordinary and graded spheres have been
reviewed in the recent literature~\cite{Lan99, Lan00, Lan01}. The
noncommutative generalization seems to be folklore.

There are two cases of particular interest.  The simplest case is when
all the constraints could be referred to as commutation relations in
the sense that they all vanish in the commutative limit. The dimension
of $V$ is equal to $n$ and the projection of $\c{A}^n$ onto
$\Omega^1(V)$ is an invertible element of $M_n(\c{A})$, a projector of
maximal rank.  Let $\t{\theta}^i =dx^i$ be a basis of
$\Omega^1(\t{\c{\c{A}}})$.  Then $\Omega^1(\c{A})$ is defined by a
projector $e \in M_n(\t{\c{A}})$ of rank $n$ with
$$
\theta^a = e^a_i dx^i
$$
a set of generators of $\Omega^1(\c{A})$.  The projection
$\pi$ has an extension to all of $\Omega^*(\c{A})$ and we can identify
$e$ as a left inverse of $\pi$.  We shall suppose that there exist $n$
matrices $\gamma^a \in M_n(\c{A})$ which satisfy the relations
\begin{equation}
\gamma^a \gamma^b = P^{ab}{}_{cd} \gamma^c \gamma^d + g^{ab}.  \label{Clif}
\end{equation}
This is a slight generalization of a proposal by Pusz and
Woronowicz\cite{PusWor89}.  The $P$ determines~\cite{DimMad96} the
structure of the algebra of forms.  The algebra generated by the
$\gamma^a$ is the generalized Clifford algebra and it bears the same
relation to the exterior algebra as its classical counterpart. The
correspondence between the Clifford and exterior algebras is only
transparent at the level of 1-forms. It is given by the module map
generated by $\gamma^a \mapsto \theta^a$. The extension to the entire
Clifford algebra is given, for example, by the map
$$
\gamma^a \gamma^b = P^{ab}{}_{cd} \gamma^c \gamma^d + g^{ab}
\mapsto P^{ab}{}_{cd} \theta^c \theta^d =  \theta^a \theta^b
$$
and consists in dropping the `symmetric' part. For this to be well
defined the $P$ would have to satisfy the braid equation. 

To every element $\imath \in M_n(\c{A})$ we associate an element $e$
defined as
\begin{equation}
e = \frac 12 (1 - \imath).                                   \label{e-i}
\end{equation}
Then $e$ is hermitian if and only if $\imath$ is hermitian and $e$
is an idempotent if and only if $\imath$ is an involution. In
particular we consider
$$
\imath = \rho^{-1}\lambda_a \gamma^a .
$$
The product here is a tensor product and $\rho \in \b{C}$. The $\imath$
will be hermitian if
$$
(\gamma^a)^* = \gamma^a, \qquad
(\rho^{-1} \lambda_a)^* = \rho^{-1} \lambda_a.
$$
The differential calculus is real if and only if the $\lambda_a$ are
antihermitian. Using~(\ref{Clif}) we can write the square of $\imath$ as
$$
\rho^2 \imath^2 = \lambda_a \lambda_b g^{ab} +
\lambda_a \lambda_b P^{ab}{}_{cd} \gamma^c \gamma^d.
$$
It is an involution if the conditions
\begin{equation}
g^{ab}\lambda_a \lambda_b = \rho^2, \qquad
P^{ab}{}_{cd}\lambda_a \lambda_b = 0                        \label{main}
\end{equation}
are satisfied.  This is the case~\cite{FadResTak89} for the sequence
of algebras $\b{R}^n_q$ for example.  In these cases the condition on
the norm of $\lambda$ is automatically satisfied and so $e$ is a
projector if and only if the commutation relations are satisfied.
This can be stated as the proposition that the element $e\in
M_n(\b{R}^n_q)$ defined in (\ref{e-i}) is an hermitian idempotent if
and only if all the relations which define $\b{R}^n_q$ are satisfied.
One notes that in these examples the `holonomic' basis does not
commute with the algebra but the basis $\theta^a$ does. sone could
more properly state that the holonomic basis $R$-commutes and the
basis $\theta^a$ does not $R$-commute. In the commutative limit $e$
tends to the identity, the unit element that is of the $SO(n)$ group.
From~(\ref{defdiff}) it follows that
$$
[e,f] = - \frac 12 [\imath, f] = \frac{1}{2}\rho^{-1} \gamma^a e_a f.
$$
This formula is to be compared with the Formula~(\ref{extra}) for the
differential. The $\gamma$-matrices are so defined that
$$
\textstyle{\det_q} (\gamma^a \gamma^b) = g^{ab}.
$$
One concludes then that $(\det_q \imath)^2 = 1$ and that one can
choose the normalization so that
$$
\textstyle{\det_q} (\imath) = 1.
$$

The paradigms are the `quantum' deformation $\b{R}^n_q$ of
euclidean space introduced~\cite{FadResTak89} by Faddeev \etal The
center is generated by a single element $r$ which one can think of as
a radius. The quotient $S^n_q$ of $\b{R}^n_q$ by the ideal generated
by the central element is known as a quantum sphere.  An element of
$M_n(\b{R}^n_q)$ has been found~\cite{LanMad01} which is an hermitian
idempotent if and only if all the relations which define
$\b{S}^{n-1}_q$ are satisfied.

\setlength{\parskip}{5pt}
\providecommand{\href}[2]{#2}\begingroup\raggedright\endgroup
%\bibliographystyle{utphys}
%\bibliography{abbrev,refgen,refmad,proceedings}
%\bibliography{/disk8/users/madore/latex/bibtex/abbrev,refgen,refmad,proceedings}

\end{document}